 \documentclass[final,5p,times,twocolumn,authoryear]{elsarticle}

\usepackage{amssymb}
\usepackage{lipsum}

\journal{Nuclear Inst. and Methods in Physics Research, A}

\begin{document}

\begin{frontmatter}



\title{From a Network to a Networking: The Evolution of the Latin American Giant Observatory}


\author[UNAB,UIS]{C. Sarmiento-Cano}
\affiliation[UNAB]{
    organization={Departamento de Ciencias Básicas, Universidad Autónoma de Bucaramanga},
    city={Bucaramanga},
    postcode={680002}, 
    state={Santander},
    country={Colombia}
    }
\affiliation[UIS]{
    organization={Escuela de Física, Universidad Industrial de Santander},
    city={Bucaramanga},
    postcode={680002}, 
    state={Santander},
    country={Colombia}    
    }   
\author[Piensa]{H. Asorey}
\affiliation[Piensa]{
    organization={piensas.xyz, Las Rozas Innova},
    addressline={Las Rozas de Madrid}, 
    city={Madrid},
    postcode={28232}, 
    country={Spain}
    }
\author[ESPOCH]{M. Audelo}
\affiliation[ESPOCH]{
    organization={Facultad de Mecánica, Escuela Superior Politécnica de Chimborazo},
    city={Riobamba},
    postcode={EC060155}, 
    country={Ecuador}
    }
\author[UniCamp]{A. Campos Fauth}
\affiliation[UniCamp]{
    organization={Instituto de Física Gleb Wataghin, Universidade Estadual de Campinas},
    city={Campinas},
    postcode={13083-859}, 
    state={SP},
    country={Brazil}
    }
\author[USFQ]{D. Cazar-Ramírez}
\affiliation[USFQ]{organization={Colegio de Ciencias e Ingenierías, Universidad San Francisco de Quito USFQ},
             city={Ecuador},
             postcode={170901},
             country={Ecuador}} 
\author[IAA,IAFE,DF]{A.M. Gulisano}
\affiliation[IAA]{
    organization={Instituto Antártico Argentino, Universidad Nacional de General San Martín},
    city={San Martín},
    postcode={B1650}, 
    state={Provincia de Buenos Aires},
    country={Argentina}
    }
    
\affiliation[IAFE]{
    organization={Instituto de Astronomía y Física del Espacio, Universidad de Buenos Aires},
    city={Buenos Aires},
    postcode={C1428EGA}, 
    state={Provincia de Buenos Aires},
    country={Argentina}
    } 
\affiliation[DF]{
    organization={Departamento de Física, Universidad de Buenos Aires},
    city={Buenos Aires},
    postcode={ C1428}, 
    state={Provincia de Buenos Aires},
    country={Argentina}
    }        
\author[UCV]{J.A. López-Rodríguez}
\affiliation[UCV]{organization={Escuela de Física, Universidad Central de Venezuela},
             city={Caracas},
             postcode={1020},
             state={Distrito Capital},
             country={Venezuela}}
\author[CIEMAT]{R. Mayo-García}
\affiliation[CIEMAT]{
    organization={Departamento de Tecnología, Centro de Investigaciones Energéticas, Medioambientales y Tecnológicas},
    city={Madrid},
    postcode={28040}, 
    state={Madrid},
    country={Spain}
    }
\author[ARRN,FIUNA]{J. Molina}
\affiliation[ARRN]{
    organization={Autoridad Reguladora Radiológica y Nuclear},
    city={Fernando de la Mora},
    postcode={2300}, 
    state={Departamento Central},
    country={Paraguay}
    }
\affiliation[FIUNA]{
    organization={Facultad de Ingeniería, Universidad Nacional de Asunción},
    city={Fernando de la Mora},
    postcode={2300}, 
    state={Departamento Central},
    country={Paraguay}
    }
\author[CONIDA]{L. Otiniano}
\affiliation[CONIDA]{
    organization={Comisión Nacional de Investigación y Desarrollo Aeroespacial},
    city={Lima},
    postcode={15046}, 
    country={Perú}
    }
\author[USAC]{J.R. Sacahui}
\affiliation[USAC]{
    organization={Escuela de Ciencias Fisicas y Matematicas, Universidad de San Carlos},
    city={Guatemala},
    postcode={01012}, 
    country={Guatemala}
    }
\author[ARRN,FIUNA]{G. Secchia-González}
\author[CAB]{I. Sidelnik}
\affiliation[CAB]{
    organization={Departamento de Física de Neutrones, Centro Atómico Bariloche},
    city={San Carlos de Bariloche},
    postcode={R8402AGP}, 
    state={Rio Negro},
    country={Argentina}
    }
\author[UIS,ULA]{L.A. Núñez}
\affiliation[ULA]{
    organization={Departamento de Física, Universidad de Los Andes},
    city={Mérida},
    postcode={5101}, 
    state={Mérida},
    country={Venezuela}    
    }

\begin{abstract}
The Latin American Giant Observatory (LAGO) is a collaborative initiative that deploys a network of low-cost, autonomous Water Cherenkov Detectors across Latin America and Spain. Initially focused on detecting gamma-ray bursts at high-altitude sites, LAGO has evolved into a multidisciplinary forum for astroparticle physics, space weather studies, and environmental monitoring. Its detectors operate from sea level to over 4300 meters above sea level (m a.s.l.) in diverse geomagnetic and atmospheric conditions. The ARTI-MEIGA simulation framework is a key development that models the entire cosmic-ray interaction chain, enabling site-specific simulations to be integrated into FAIR-compliant workflows. LAGO also plays a significant role in regional education and training through partnerships with ERASMUS+ projects, positioning itself as a hub for research capacity building. New contributions emerging from the collaboration include volcano muography, neutron hydrometry for precision agriculture, and space weather monitoring in the South Atlantic Magnetic Anomaly. LAGO demonstrates how Cherenkov-based detection and open science can drive scientific discovery and practical innovation
\end{abstract}




\end{frontmatter}



\section{The LAGO Collaboration}
The Latin American Giant Observatory (LAGO) originated as a spin-off inspired by the Pierre Auger Observatory experience and was formally launched in the mid-2000s with a focused scientific goal: deploy autonomous Water Cherenkov Detectors (WCDs), especially at high-altitude Andean sites, to search for the high-energy component of gamma-ray bursts \citep{sidelnik_sites_2015}. Over time, that initial ``install detectors and measure'' approach expanded into a geographically and operationally diverse observatory, with sites spanning from Mexico to Antarctica (and Spain). This diversity has become a defining scientific asset because it enables simultaneous observations across very different atmospheric depths and geomagnetic conditions. Thus, LAGO developed a space weather program, which is today one of the main objectives of the collaboration \citep{asorey_lago_2015}.

From the outset, LAGO has been defined not only by its technical goals but also by its educational and institutional mission. Over time, the collaboration has become a regional hub for training and capacity building, leveraging international partnerships and structured learning activities to provide students and early-career researchers with hands-on experience with instrumentation, simulation tools, and open-data practices.

\section{The network}
\label{Network}
The LAGO detector network comprises autonomous WCDs installed covering a wide range of altitudes and geomagnetic rigidity. These detectors currently operate from sea level to 4300 m a.s.l. at sites such as Chacaltaya (Bolivia) and Chimborazo (Ecuador). 

The LAGO WCD consists of a sealed, light-tight container filled with water and equipped with optical detectors, typically an 8''-9'' photomultiplier tube (PMT). These detectors are in close contact with the water volume to register Cherenkov radiation. As Cherenkov photons propagate through the water volume and may reach the optical detector, resulting in a variable number of photoelectrons, which generate a current that is amplified in the corresponding stages of the optical detector. In the LAGO detector network, an updated acquisition electronics system is currently being deployed. This update enables higher-resolution temporal evolution of the measured pulses by the WCD and of the obtained energy spectra\citep{Arnaldi2020TheND}.

Designed using low-cost, open-source hardware, LAGO WCD stations can operate continuously with minimal maintenance. The collaboration is advancing the concept of a \emph{WCD as an appliance}: a detector conceived not as a one-off laboratory setup, but as a standardized, self-contained instrument that can be deployed, configured, and operated in a largely uniform way across the network. In practice, the appliance model aims to provide a plug-and-play station: a reproducible hardware and firmware stack with well-defined interfaces for power, timing, data acquisition, and communications, enabling new sites to be commissioned quickly and existing sites to be maintained and upgraded with minimal local intervention. For calibration, LAGO device a new method \citep{otiniano_measurement_2023} that detects Michel electrons from muon decays, eliminating the need for external instrumentation. This method has proven stable across diverse environmental and geomagnetic conditions complementing the standard VEM scheme.

\section{The networking}
\label{Networking}
Today, LAGO's research extends its focus to building a scientific network that links researchers across Latin America and Spain through real-time, distributed collaboration. It retained its astroparticle physics roots while developing strong programs in space weather and atmospheric radiation, and it opened pathways to applied research, including muography, radiation dose studies, and environmental monitoring. 

\subsection{ARTI-MEIGA framework}
The LAGO space weather program drove the development of a precise computational toolkit capable of simulating both extended atmospheric showers and the detector response at ground level. 

Today, the ARTI-MEIGA framework \citep{sarmiento_arti_2022,taboada_meiga_2022} is a publicly available, highly configurable simulation chain designed to compute the signals produced by secondary particles resulting from the interaction of primary cosmic rays with the atmosphere. It can model scenarios ranging from single primaries and selected particle populations to the integrated response to the full galactic cosmic-ray flux. In addition, ARTI can be used to estimate site-dependent ground-level fluxes and detector signals associated with transient astrophysical phenomena at any location worldwide.

The ARTI-MEIGA framework has supported a wide range of studies and applications, including radiation dose assessments for commercial flight crews \citep{asorey_acorde_2023}, astroparticle physics analyses \citep{sidelnik_capability_2023}, muography for volcanic \citep{vesga-ramirez_muon_2020} and industrial monitoring \citep{MartinezRivero_MuonImaging_2025}, and emerging smart-agriculture use cases \citep{betancourt_enhanced_2025}. 

ARTI is distributed through the LAGO GitHub repository~\citep{artiGitHub} and can be executed across a wide range of computing resources, from personal laptops and desktops to high-performance computing clusters. More recently, it has been adapted to run in cloud environments via \texttt{OneDataSim}~\citep{rubio_novel_2021}, enabling reproducible deployments on the European Open Science Cloud (EOSC) and other federated or public clouds. This cloud-based service is also available through the EOSC Marketplace~\citep{marketplace}.

\subsection{Blockchain, edge computing, digital twins and all that}
LAGO’s edge-computing and blockchain efforts \citep{MierBello_CARLA_2025, MartinezMendez_CARLA_2025} demonstrate its evolution into a digital observatory. On the edge, WCD stations are treated as ``appliances'' capable of local preprocessing and data transfer, reducing dependence on continuous high-bandwidth links. To strengthen trust in a geographically distributed setting, a permissioned blockchain approach stores hashes and metadata on-chain while keeping large datasets off-chain in existing repositories.

LAGO is also implementing AI-enhanced data analysis at the edge and seamless integration into FAIR-compliant workflows in the cloud, ensuring that all data is Findable, Accessible, Interoperable, and Reusable \citep{torres_enhanced_2024}.

\subsection{The partnership with ERASMUS+ CBHE projects}
LAGO’s training vocation has always gone beyond the technical: from the beginning, the collaboration was envisioned as a regional, distributed ``school of practice'' in astroparticle physics, where learning happens through building and operating detectors, analyzing real data, and sharing tools within the community. In this spirit, LAGO contributes to ERASMUS capacity-building initiatives such as LA-CoNGA and EL-BONGÓ by offering a genuine research infrastructure of distributed detectors, live data streams, and integrated analysis and simulation pipelines that serve as a hands-on training environment.

LA-CoNGA physics\footnote{LA-CoNGA physics for Latin American alliance for Capacity buildiNG in Advanced physics https://laconga.redclara.net/} is an Erasmus+ Capacity Building in Higher Education project focused on modernizing advanced physics education in the Andean region and building shared learning environments across universities \citep{PenarodriguezNunez_CONGA_2022}. EL-BONGÓ physics \footnote{EL-BONGÓ physics for E-Latin American huB for OpeN Growing cOmmunities in Physics https://elbongo.redclara.net/} scales and extends to develop four research and learning communities \citep{Caicedo_Virtual_2017}. EL-BONGÓ physics adds a FABLab network training in digital fabrication, building ``do-it-yourself" skills to develop and maintain low-cost scientific instruments.

\subsection{Extended capability of WCD}
\subsubsection{Neutron detection}
Recent studies \citep{Sidelnik_enhancing_2020A,Sidelnik_Neutron_2020B,Sidelnik_500MeV_2020C} show that adding small amounts of sodium chloride (NaCl) to the detector water can significantly improve neutron sensitivity. The enhancement comes from thermal neutrons capturing on chlorine nuclei, which then emit cascades of gamma-ray signals. These gammas produce secondary electrons in water, which, in turn, generate additional Cherenkov light, thereby amplifying the measured signal. Salt-doped WCDs could provide a viable complement, or even an alternative, to traditional $^3{\rm He}$-based neutron monitors in applications ranging from environmental radiation surveillance to the detection of special nuclear materials.

\subsubsection{From PMTs to SiPMs}
For LAGO, transitioning from conventional PMTs to silicon photomultipliers (SiPMs) can significantly strengthen the case for the Water Cherenkov Detector as an appliance. Traditional PMTs provide a large sensitive area, but they come with practical costs. PMTs typically require high-voltage systems operating at kilovolts, specialized insulation and connectors, and careful handling because they are fragile vacuum devices. These needs increase installation complexity, raise safety concerns, and make maintenance more challenging in limited-infrastructure settings. By contrast, SiPMs are compact, solid-state sensors that operate at low voltage (tens of volts), enabling simpler and safer power distribution, easier electronics integration, and improved mechanical robustness during transport and field deployment. In an appliance mindset, this translates directly into stations that are more uniform, easier to replicate across institutions, and more resilient under variable local conditions.

A key obstacle is sensor area: single SiPMs are much smaller than the PMTs traditionally used in WCDs. In this context, C-ARAPUCA illustrates a viable alternative by combining SiPMs with an optical photon-collection and trapping scheme (using wavelength shifting and reflective confinement) that effectively increases light collection without relying on a large PMT \citep{fauth_ARAPUCA_2024}. In this way, C-ARAPUCA-like concepts can preserve practical signal levels while keeping the station low-power and easy to standardize—exactly what an appliance requires.

\subsection{Application with social impacts}
\subsubsection{Soil moisture and smart agriculture}
This chloride-assisted approach extends the capability of WCDs toward soil-moisture sensing and smart-agriculture applications. Because NaCl is inexpensive, widely available, and non-toxic at the concentrations of interest, it offers a practical route to boosting thermal-neutron detection \citep{betancourt_enhanced_2025}. Moreover, the large effective footprint of WCD stations (up to hectare scale) and their robustness make them attractive complements to existing cosmic-ray neutron sensing networks for hydrology and precision agriculture. Future field campaigns will validate the simulated performance gains against in situ moisture probes, evaluate the long-term optical stability of brine solutions, and test combined epithermal and thermal measurement strategies. Integrating low-power electronics and edge processing will further support autonomous, long-term deployments.

\subsubsection{Muography}
Muon radiography uses naturally produced atmospheric muons as a non-invasive probe to map density variations inside large structures. For LAGO, this is a natural extension of its core expertise: the same understanding of cosmic-ray secondaries and detector response can be redirected toward practical problems—without relying on artificial radiation sources. As a result, muography broadens LAGO's impact beyond astroparticle physics into geophysics, civil protection, and industrial inspection.

MuTe (for Muon Telescope and also named after a local soup from Santander, Colombia) exemplifies this transition. It is a hybrid instrument that combines two scintillator panels for muon tracking with a Water Cherenkov Detector used for background rejection and energy-deposition measurements \citep{pena_design_2020design}. It directly reuses the WCD know-how developed across LAGO and adapts it to demanding field conditions. Its emphasis on low power consumption, robustness, and portability reflects the same "appliance mindset" that LAGO promotes for distributed instrumentation: systems designed to be reliable, reproducible, and deployable in remote locations.

The 2025 hydrotreatment-reactor study \citep{MartinezRivero_MuonImaging_2025} illustrates the next step in maturity. LAGO members translate muography into industrial diagnostics by designing and validating a muon-imaging system to monitor catalyst beds in high-pressure reactors. The work integrates the ARTI-MEIGA toolchain and validates detector performance through controlled attenuation measurements before advancing to realistic three-dimensional reactor models. This progression from concept and calibration to simulation-backed deployment highlights how muography strengthens LAGO's evolution into a network of people innovating with cosmic rays.

\subsubsection{The South Atlantic Magnetic Anomaly cosmic ray flux}
The South Atlantic Magnetic Anomaly (SAMA/SAA) is characterized by an unusually weak geomagnetic field, which reduces Earth's natural shielding and allows charged particles to penetrate to lower altitudes more easily than in most regions. It constitutes a unique natural laboratory for studying how solar activity and geomagnetic disturbances translate into measurable changes in ground-level secondary-particle fluxes and atmospheric conditions over populated areas.

The Paraguay study shows that a low-cost, compact muon detector can measure the muon flux with sufficient stability to detect Forbush decreases and reveal statistically significant correlations with geomagnetic activity during events in May and October 2024 \citep{molina_measurements_2025}. 

Beyond instrumentation, the broader regional perspective underscores why SAA-focused space-weather studies matter beyond physics. During intense solar events, enhanced particle precipitation over the anomaly can increase atmospheric ionization, activating chemical pathways ($NO_x/HO_x$) that contribute to ozone depletion and potentially modify surface UV exposure. Linking geomagnetic vulnerability to particle flux, atmospheric chemistry, and environmental effects frames SAA/SAMA research as an inherently interdisciplinary effort—one in which measurements are a key component of a larger climate and health-relevant system.

\section*{Acknowledgements}
The authors acknowledge co-funding from the Programa Iberoamericano de Ciencia y Tecnología para el Desarrollo (CYTED) through the LAGO-INDICA network (Project 524RT0159-LAGO-INDICA: Infraestructura digital de ciencia abierta). A.M.G. and IS are members of the Carrera del Investigador Científico of CONICET. This work was supported by the Argentinean grants PICT 2019-02754 (FONCyT-ANPCyT) and UBACyT-20020190100247BA (UBA). AI technology was used to proofread and polish this manuscript. (OpenAI, 2026). After using this tool, the authors reviewed and edited the content as needed and take full responsibility for the content of the published article.







\end{document}